\documentclass[
    ,final            % use final for the camera ready runs
    ,sort&compress                 % add further options here if necessary
  ]
  {aipproc}

\layoutstyle{8x11single}

\usepackage[normalem]{ulem} % \sout{old text} for strikeout
\usepackage[dvips]{color} % For in-text comments and additions in color
\begin{document}

\title{Pion photoproduction in a dynamical coupled-channel model}



\author{F.~Huang}{
  address={Department of Physics and Astronomy, University of Georgia, Athens, GA 30602, USA}
}

\author{M.~D\"oring}{
 address={Institut f{\"u}r Kernphysik and J\"ulich Center for Hadron Physics,
          Forschungszentrum J{\"u}lich, 52425 J{\"u}lich, Germany} }

\author{H.~Haberzettl}{
  address={Center for Nuclear Studies, Department of Physics, The George Washington
            University, Washington, DC 20052, USA}
  ,altaddress={Institut f{\"u}r Kernphysik and J\"ulich Center for Hadron Physics,
               Forschungszentrum J{\"u}lich, 52425 J{\"u}lich, Germany} }

\author{S.~Krewald}{
 address={Institut f{\"u}r Kernphysik and J\"ulich Center for Hadron Physics,
          Forschungszentrum J{\"u}lich, 52425 J{\"u}lich, Germany}
 ,altaddress={Institute for Advanced Simulations,
              Forschungszentrum J\"ulich, 52425 J\"ulich, Germany} }

\author{K.~Nakayama}{
  address={Department of Physics and Astronomy, University of Georgia, Athens, GA 30602, USA}
 ,altaddress={Institut f{\"u}r Kernphysik and J\"ulich Center for
Hadron Physics,
Forschungszentrum J{\"u}lich, 52425 J{\"u}lich, Germany} % additional visiting address
}

\begin{abstract}
Pion photoproduction reactions are investigated in a dynamical
coupled-channel approach based on the J\"ulich $\pi N$ model, which presently
includes the hadronic $\pi N$ and $\eta N$ stable channels as well as the
$\pi\Delta$, $\sigma N$ and $\rho N$ effective channels. This model has
been quite successful in the description of $\pi N \to \pi N$ scattering for
center-of-mass energies up to $1.9$ GeV. The full pion photoproduction
amplitude is constructed to satisfy the generalized Ward-Takahashi identity and
hence, it is fully gauge invariant. The calculated differential cross
sections and photon spin asymmetries up to 1.65 GeV center-of-mass energy for
the reactions $\gamma p\to \pi^+n$, $\gamma p\to \pi^0p$ and $\gamma n\to
\pi^-p$ are in good agreement with the experimental data.
\end{abstract}

\maketitle

%%%%%%%%%%%%%%%%%%%%%%%%%%%%%%%%%%%%%%%%%%%%
%% MAINMATTER
%%%%%%%%%%%%%%%%%%%%%%%%%%%%%%%%%%%%%%%%%%%%

\section{Introduction}

In the past few years, intensive experimental efforts have taken
place at major laboratories around the world to study the
baryon spectra using the electromagnetic probes.  On the theoretical
side, since the baryon resonances are unstable and couple strongly
to baryon-meson continuum states, coupled-channel models are needed
to analyze the experimental data and to interpret the
extracted resonance parameters. For our present study, we employ the
well-known  J\"ulich $\pi N$ model \cite{Schutz98,Krehl00,Gasparyan03} that is
based on time-ordered perturbation theory. It is a coupled-channel
meson-exchange model which includes the $\pi N$ and $\eta N$ stable channels as well as the $\pi\Delta$,
$\sigma N$ and $\rho N$ effective channels accounting for the resonant part of the $\pi\pi N$ channel. The interaction kernel corresponding to the $t$- and
$u$-channel diagrams is constructed based on the chiral Lagrangians of Wess and
Zumino \cite{Wess67}, supplemented by additional terms for the coupling of
$\Delta$, $\omega$, $\eta$, $a_0$ and $\sigma$ \cite{Krehl00,Gasparyan03}. For
the $s$-channel diagrams, apart from the bare nucleon dressed
by the coupling to the $\pi N$ continuum state to reproduce the physical
nucleon, the interaction kernel includes eight genuine bare resonances,
namely $S_{11}(1535)$, $S_{11}(1650)$, $S_{31}(1620)$, $P_{31}(1910)$,
$P_{13}(1720)$, $D_{13}(1520)$, $P_{33}(1232)$ and $D_{33}(1700)$. The bare
genuine resonances get their dressed masses and widths from the re-scattering
of the baryon-meson continuum states; the $P_{11}(1440)$ (Roper)
resonance appears as a dynamically generated resonance due to the strong
coupling of the $\pi N$ and $\sigma N$ channels. This hadronic model has been
quite successful in reproducing the $\pi N$ partial-wave amplitudes up
to the center-of-mass energy of $W=1.9$ GeV and for the total angular momentum
states with $J=1/2$ and $3/2$.

In Ref.~\cite{Haberzettl06} a dynamical coupled-channel model for pseudoscalar
meson photoproduction based on a field-theoretical approach of
Haberzettl \cite{Haberzettl97} has been introduced in conjunction with the
J\"ulich hadronic coupled-channel model. This photoproduction model is
distinguished from the majority of the existing dynamical models by the fact
that, in addition to satisfying unitarity and analyticity as a matter of
course, it also satisfies the full gauge-invariance condition
dictated by the generalized Ward-Takahashi identity. By
contrast, the vast majority of existing dynamical models satisfy
only current conservation but not gauge invariance. In
Ref.~\cite{Haberzettl06}, a preliminary feasibility study of this
field-theory approach was presented. In the present
work, we carry out an extended and more quantitative calculation of the neutral
and charged pion photoproduction reactions within this approach using the
J\"ulich $\pi N$ model \cite{Gasparyan03}. We
calculate cross sections as well as the beam asymmetries up to $W=1.65$
GeV.

The full single-meson photoproduction amplitude, $M^\mu$, is obtained
by attaching a photon everywhere into the dressed $NN\pi$ vertex, which gives
\cite{Haberzettl06}
\begin{equation}
M^\mu = M^\mu_s + M^\mu_u + M^\mu_t + M^\mu_{\rm int},
\label{eq:Mmu}
\end{equation}
where the first three terms describe the coupling of the photon to the external
legs of the dressed $NN\pi$ vertex, with the subscripts $s$, $u$ and
$t$ referring to the appropriate Mandelstam variables of their respective
intermediate hadrons. The last term $M^\mu_{\rm int}$ is interaction current,
where the photon couples inside the $NN\pi$ vertex. In our work it is chosen as
\cite{Haberzettl06}
\begin{equation}
M^\mu_{\rm int} = M^\mu_c + T^\mu + T^{NP} G_0 \left[\left(M^\mu_u +
M^\mu_t\right)_T + T^\mu\right], \label{eq:Mint2}
\end{equation}
where $\left(M^\mu_u + M^\mu_t\right)_T$ denotes the transverse part of
$\left(M^\mu_u + M^\mu_t\right)$; $T^\mu$ stands for an undetermined transverse
contact current taken to be zero in the present work for
simplicity; $T^{NP}$ is the non-polar part of the hadronic scattering
amplitude; $G_0$ describes the free propagation of the intermediate
meson-baryon two-body system. $M^\mu_c$ is the generalized contact current as
specified in Ref.~\cite{Haberzettl06}; it accounts for the complicated part of
the interaction current which cannot be treated explicitly. The amplitude
$M^\mu_s$ in Eq.~(1) is given by
\begin{equation}
M^\mu_s = \Gamma_{RN\pi} \,S\, \Gamma^\dag_{RN\gamma},
\end{equation}
where $\Gamma_{RN\pi}$ stands for the dressed $R\to N\pi$ vertex with $R$
denoting either the nucleon or baryon resonance, $S$ denotes the dressed
baryon propagator. The dressed hadronic vertices and propagators are determined
by the hadron dynamics of the J\"ulich $\pi N$ coupled-channel model.
$\Gamma^\dag_{RN\gamma}$ is the full dressed $N\gamma \to R$ vertex as derived
in Ref.~\cite{Haberzettl06}.

\section{Results}

\begin{figure}[t!]
\includegraphics[width=0.65\textheight]{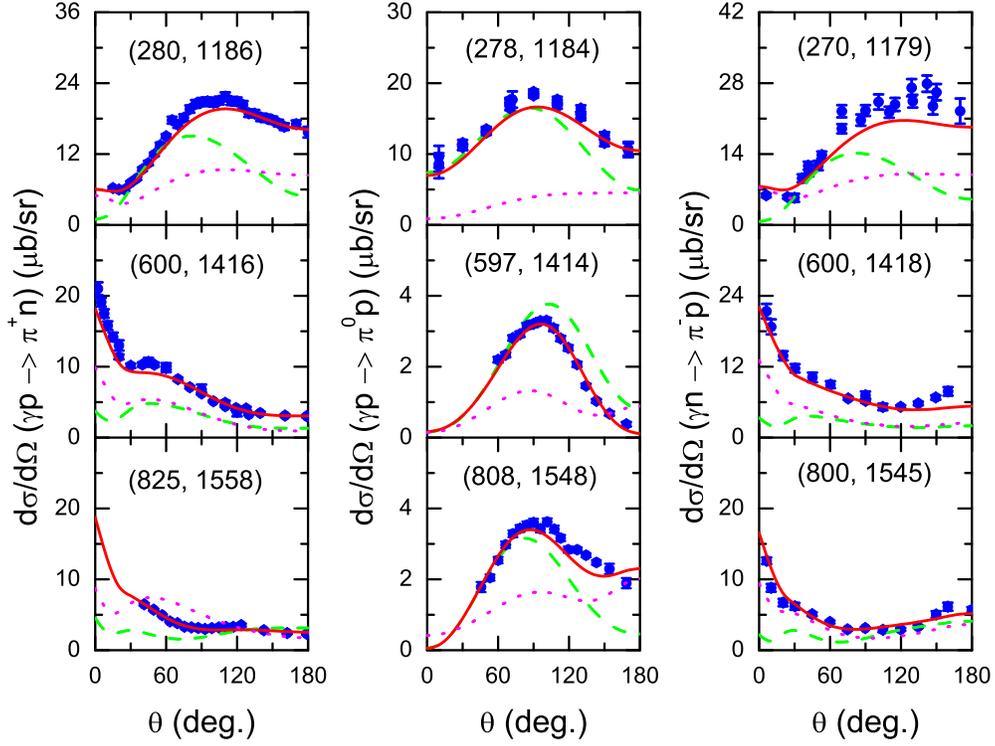}
\caption{\label{fig:dsdo}Differential cross sections for $\gamma p
\to \pi^+ n$, $\gamma p\to \pi^0 p$ and $\gamma n \to \pi^- p$. The
solid curves show the results from the full calculation. The dotted
and the dashed curves are obtained by respectively switching off the loop integral
or the contact current $M^\mu_c$. Data are
from
%Refs.~\cite{Dugger09,Ahrens04,SAID,Bartalini05,Bartholomy05,Shafi04,Elsner09}.}
Ref.~\cite{SAID}.}
\end{figure}

\begin{figure}
\includegraphics[width=0.65\textheight]{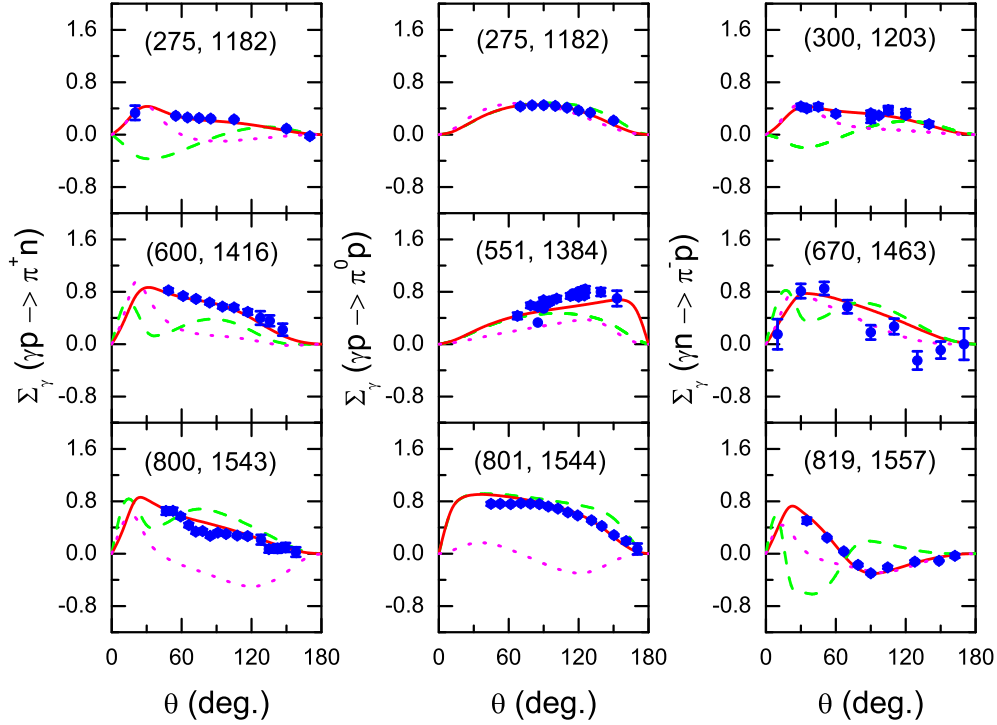}
\caption{\label{fig:sa}Photon spin asymmetries for $\gamma p \to
\pi^+ n$, $\gamma p\to \pi^0 p$ and $\gamma n \to \pi^- p$. The
solid curves show the results from the full calculation. The dotted
and the dashed curves are obtained by respectively switching off the loop integral
or the contact current $M^\mu_c$. Data are
from
%Refs.~\cite{Dugger09,Ahrens04,SAID,Bartalini05,Bartholomy05,Shafi04,Elsner09}.}
Ref.~\cite{SAID}.}
\end{figure}

\begin{figure}
\includegraphics[width=0.44\textwidth]{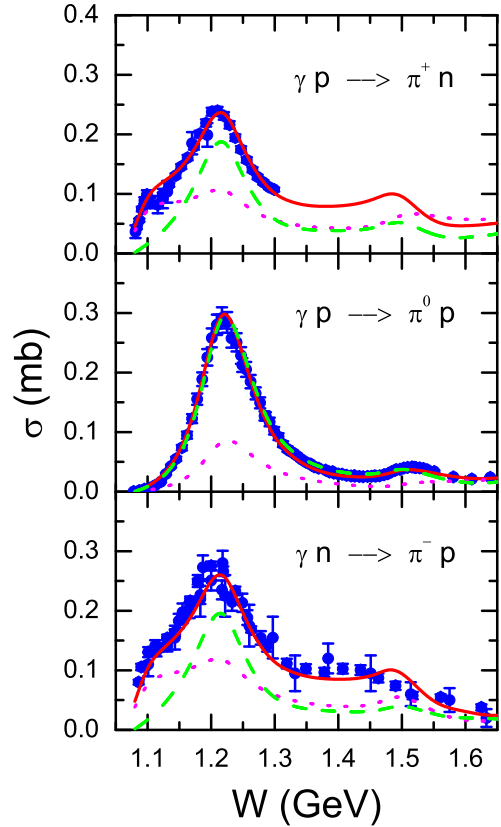}
\caption{\label{fig:sig}Total cross sections for $\gamma p \to \pi^+
n$, $\gamma p\to \pi^0 p$ and $\gamma n \to \pi^- p$. The solid
curves show the results from the full calculation. The dotted and
the dashed curves are obtained by respectively switching off the loop integral
or the contact current $M^\mu_c$. Data are taken from
Ref.~\cite{SAID} but not included in the fit.}
\end{figure}

The parameters of the present pion photoproduction model are determined by
fitting the differential cross section and the beam asymmetry data of the
charged and neutral pion photoproduction up to $W=1.65$ GeV. They are listed in
Ref.~\cite{Huang10}. The calculated differential cross
sections and beam asymmetries for $\gamma p\to \pi^+ n$, $\gamma p\to \pi^0 p$
and $\gamma n\to \pi^- p$ are shown in Figs.~\ref{fig:dsdo} and \ref{fig:sa},
respectively, together with the corresponding experimental data. More
results for differential cross sections and beam asymmetries at 13 additional
energies from the $\pi N$ threshold up to $W=1.65$ GeV can be found in
Ref.~\cite{Huang10}. Figure~\ref{fig:sig} shows the total cross
sections for the three charged pion photoproduction channels. Note that the
total cross-section data have not been included in our fit. In
Figs.~\ref{fig:dsdo}-\ref{fig:sig}, the solid curves correspond to the results
of our full calculation, while the dotted and dashed curves are obtained by
respectively switching off the loop integral and the contact current
$M^\mu_c$ in the full amplitude. One sees that our theoretical
results for both the differential cross sections and beam asymmetries
(Figs.~\ref{fig:dsdo} and \ref{fig:sa}) as well as the total cross sections
(Fig.~\ref{fig:sig}) are in good agreement with the corresponding experimental
data. Juli\'{a}-D\'{\i}az {\it et al.} \cite{Bruno08} have also performed an
investigation of the single-pion photoproduction within the EBAC
dynamical coupled-channel model. Compared to our model, they included
two additional resonances, namely $D_{15}$ and $F_{15}$;
however, the $\gamma n\to \pi^-p$ channel was not analyzed in
their work. Compared with the results of Ref.~\cite{Bruno08}, the agreement of
our results with the data is a little bit better for both the $\gamma p\to
\pi^+ n$ and $\gamma p\to \pi^0 p$ channels. Figures~\ref{fig:dsdo}-\ref{fig:sig} also show clearly that the loop integral and the contact current $M^\mu_c$ are important for reproducing the data, which means that the
coupled-channel effects \textit{and} gauge invariance are essential for
understanding the single-pion photoproduction reactions. While
the importance of coupled-channel effects was pointed out already in
Ref.~\cite{Bruno08}, we emphasize here that the majority of existing dynamical
models of photoproduction reactions are actually not gauge invariant since
their amplitudes obey current conservation only. The latter is shown in
Ref.~\cite{Haberzettl97} to be necessary but not sufficient for
maintaining full gauge invariance across all levels of a microscopic
description of the photoreaction. The importance of fulfilling the
gauge-invariance condition was also emphasized for the $NN$
bremsstrahlung reaction \cite{Nakayama09}.

Finally, we mention that the residues and pole positions of the full
photoproduction amplitude are being evaluated in our model by making an
analytic continuation of the amplitudes to complex energy plane. The results
will be published elsewhere.

%%%%%%%%%%%%%%%%%%%%%%%%%%%%%%%%%%%%%%%%%%%%%%%%
%% BACKMATTER
%%%%%%%%%%%%%%%%%%%%%%%%%%%%%%%%%%%%%%%%%%%%%%%%

\begin{theacknowledgments}
This work is supported by the FFE grant No. 41788390 (COSY-058). The
authors acknowledge the Research Computing Center at the University
of Georgia and the J\"ulich Supercomputing Center at
Forschungszentrum J\"ulich for providing computing resources that
have contributed to the research results reported within this
proceeding.
\end{theacknowledgments}

\end{document}